\documentclass[prb,twocolumn,groupeaddress]{revtex4-1}

\usepackage{amsmath}
\usepackage{amssymb}
\usepackage{xspace}
\usepackage{graphicx}
\usepackage{grffile}
\usepackage{nicefrac}
\usepackage{array}
\usepackage{color}

\graphicspath{{figs/}}
\newcolumntype{L}[1]{>{\raggedright\let\newline\\\arraybackslash\hspace{0pt}}m{#1}}
\newcolumntype{C}[1]{>{\centering\let\newline\\\arraybackslash\hspace{0pt}}m{#1}}
\newcolumntype{R}[1]{>{\raggedleft\let\newline\\\arraybackslash\hspace{0pt}}m{#1}}

\def\bk{{\bf k}}

\begin{document}

\title{Doping-dependent character and possible magnetic ordering of NdNiO$_2$}
\author{Frank Lechermann}
\affiliation{European XFEL, Holzkoppel 4, 22869 Schenefeld, Germany}
\affiliation{Center for Computational Quantum Physics, 
The Flatiron Institute, 162 5th Avenue, New York, NY 10010, USA}

\begin{abstract}
The novel nickelate superconductors of infinite-layer type feature 
challenging electronic pecularities in the normal-state phase diagram
with doping. Distinct many-body behavior and different dispersion regimes of the 
entangled $\{$Ni-$d_{z^2}$, Ni-$d_{x^2-y^2}$$\}$ orbital sector give
rise to highly rich physics, which is here studied for the case of 
the NdNiO$_2$ system. An analysis based on advanced realistic dynamical 
mean-field theory unveils that the superconducting hole-doped region is the
meeting place of a (self-)doped Mott insulator
from the underdoped side, and a bad Hund metal from the overdoped
side. Fermi-level crossing of the Ni-$d_{z^2}$ flat-band ties both 
regimes together to form a singular arena for unconventional 
superconductivity. We furthermore shed light on the intriguing 
problem of elusive magnetism in infinite-layer nickelates.
Antiferromagnetic (AFM) order with small Ni moments is shown to be 
a vital competitor at low temperature. At stoichiometry, C-AFM order
with ferromagnetic spin-alignment along the $c$-axis benefits from a 
conceivable coexistence with Kondo(-lattice) screening. 
\end{abstract}

\maketitle

\section{Introduction} 
Nickelate superconductivity in thin films of Sr-doped NdNiO$_2$ below a
critical temperature $T_{\rm c}\sim 10$\,K entered the scene in summer 
2019~\cite{li19}. Recently, the same physics has also been identified in 
Sr-doped PrNiO$_2$ films~\cite{osa20-1,osa20-2}. Because of the 
sophisticated sample preparation and the so far lack of bulk single crystals, 
experimental progress on these Ni($3d^9$)-based infinite-layer systems 
is challenging. 

Several findings still enable further insight into a very rich phyenomenology.
Ordered magnetism is hard to measure in the thin-film geometry, but neutron scattering
on polycrystalline bulk samples provides no clear evidence for long-range 
antiferromagnetic (AFM) order in NdNiO$_2$~\cite{hay03,wang20}. However, recent 
nuclear-magnetic-resonance (NMR) experiments~\cite{cui20} on similar samples suggest
quasistatic AFM ordering for $T<40$\,K. This, together with a resistivity 
upturn~\cite{li19} below $T\sim 70$\,K, points to intriguing low-energy physics even far 
from the superconducting region. That region is sandwiched between weakly-insulating doping 
regimes, in the range $0.125\lesssim x\lesssim 0.25$ of the phase diagram for 
hole doping $x$ of Sr kind~\cite{li20,zen20}. Doping-dependent Hall 
data~\cite{li20,zen20} point to a two-band scenario with $x$. First spectroscopic 
results reveal hybridization of 
Ni$(3d)$ states with other delocalized states and reduced Ni-O intermixing compared 
to NiO and LaNiO$_3$~\cite{hep20}. Furthermore for finite $x$, substantial Ni$(3d^8)$ 
character is unveiled~\cite{goo21,ros20}, in contrast to the dominant 
Cu$(3d^9\underline{L})$, i.e. 
Cu$(3d^9)$ with ligand hole on oxygen, weight in high-$T_{\rm c}$ cuprates. This 
Ni$(3d^8)$ signature is either attributed to multi-orbital physics~\cite{goo21} 
or to Ni($S=0$) single-orbital physics of $d_{x^2-y^2}$ kind~\cite{ros20}. Concerning
the superconducting state, a small anisotropy in the upper critical field, 
hinting toward a Pauli-limited coupling, as well as a two-symmetry gap nature are 
reported~\cite{xia20,wanli21,gu20}. 
 
From the theoretical side, assessments based on density functional theory (DFT) 
highlight the self-doping (SD) nature of an otherwise seemingly cuprate-like band 
structure~\cite{wu19,nom19,bot20,lec20-1,lec20-2}. The weakly-filled SD band originates 
from Nd$(5d)$-Ni$(3d)$ hybridization and causes a multi-sheeted Fermi surface at 
stoichiometry. 
There seems also agreement that the charge-transfer character of NdNiO$_2$ is reduced
compared to estimates in akin cuprates~\cite{jia19,lec20-1}. Further beyond DFT, two main 
theory concepts are pursued: either (decorated) single-Ni-orbital physics of 
Ni-$d_{x^2-y^2}$ type~\cite{wu19,nom19,bot20,zha20,si20,kit20,kar20-1,adh20,lan20,bee21}, 
or multiorbital Ni($3d)$ 
processes~\cite{lee04,hu19,zhav20,wer20,lec20-1,lec20-2,cha20,cho20-2,pet20,wankang20,kan20} 
are assumed key to the correlation 
phenomenology. In two previous works~\cite{lec20-1,lec20-2}, we showed that based on
calculations utilizing the combination of DFT with dynamical mean-field theory (DMFT) and 
furthermore including explicit Coulomb interactions on oxygen, the relevance of 
Ni-multiorbital degrees of freedom is indeed inevitable. Besides the indisputable 
Ni-$d_{x^2-y^2}$ orbital and its related low-energy band, the Ni-$d_{z^2}$ orbital plays 
a further crucial role. First, in mediating a Kondo(-lattice) coupling between
Ni$(3d)$ and Nd$(5d)$ at stoichiometry. And second, for establishing a 
unique flat-band scenario in the doping region of low-$T$ superconductivity. 

In the present work, the goal is to extent our previous studies and shed further light 
onto the Ni-multiorbital physics in infinite-layer nickelates. In 
Refs.~\onlinecite{lec20-1,lec20-2}, paramagnetism has been assumed and doping has 
been addressed by a supercell description~\cite{lec20-1} as well as within a 
minimal three-band model solved for a simplified Fermi-liquid setting~\cite{lec20-2}. 
Here, we provide a detailed account of the doping-dependent features of the 
correlated electronic strucuture based on the virtual-crystal approximation, and 
additionally investigate the intriguing problem of magnetic ordering in the NdNiO$_2$ 
system.

\section{Methodology\label{sec:med}}
\begin{figure}[b]
\includegraphics*[width=8.5cm]{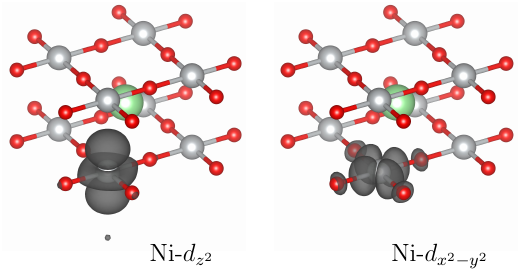}
\caption{(color online) Projected local orbitals of Ni-$d_{z^2}$ (left) and 
Ni-$d_{x^2-y^2}$ (right) type in the NdNiO$_2$ compound: 
Ni(grey), Nd(large,green), O(small,red).}
\label{fig:orb}
\end{figure}

The NdNiO$_2$ compound crystallizes with $P4/mmm$ space group, giving rise
to a four-atom unit cell (see Fig.~\ref{fig:orb} for the enlarged cell). 
The occupied Wyckoff positions amount to 1a:(0,0,0) for Ni, 1d:(0.5,0.5,0.5) for Nd as 
well as 2f:(0.5,0,0) for O. Stoichiometric lattice parameters are taken 
from experiment~\cite{li19}, reading $a=3.92$\,\AA\; and $c=3.31$\,\AA.
\subsection{DFT+sicDMFT framework} 
The charge self-consistent combination~\cite{sav01,gri12} of DFT and DMFT with the
additional incorporation of the self-interaction correction (SIC) is used to investigate
the normal-state properties of pristine and doped neodynium nickelate in the
infinite-layer structure. The realistic part of this DFT+sicDMFT scheme~\cite{lec19}
builds up on a mixed-basis pseudopotential representation~\cite{els90,lec02,mbpp_code} 
in the local density approximation (LDA). Coulomb interactions on oxygen are  
addressed beyond DFT within SIC applied to the O pseudopotential. On the transition-metal
(TM) sites, those interactions are treated within DMFT. All three contributions, i.e.
DFT, SIC and DMFT are coupled within a fully charge self-consistent framework that
is iterated until convergence of the correlated electronic structure is reached.

A detailed account of the utilized SIC construction can be found in Ref.~\onlinecite{kor10} 
and references therein. Let us here recapitulate only the main equations on 
the (pseudo-)atomic level, i.e.
\begin{align}
&(-\nabla^2+V_l+V_{\rm H}[n_{v}]+V_{\rm xc}[n_{v}]+V_{\rm cor}[n_{l}])\Psi_l^{pp}
=\epsilon_l^{pp}\Psi_l^{pp}\;,\label{eq1}\\
&V_{\rm cor}[n_{l}]=-w_l[V_{\rm H}[n_{l}]+V_{\rm xc}[n_{l}]]\;,\label{eq2}\\
&V_l^{\rm SIC}(r):=V_l(r)-\alpha\langle\Psi_l^{pp},V_{\rm cor}[n_{l}]\Psi_l^{pp}\rangle
\Psi_l^{pp}(r)\;,\label{eq3}
\end{align} 
where $\langle\cdots,\cdots\rangle$ marks the scalar product. In equation~(\ref{eq1})
the standard norm-conserving pseudopotential $V_l$ for angular-momentum quantum 
number $l$, Hartree potential $V_{\rm H}$ and exchange-correlation potential 
$V_{\rm xc}$ for the density $n_v$ of all valence electrons, as well as 
the correction potential $V_{\rm cor}$ govern the wave function
$\Psi_l^{pp}$. Thereby, $n_v=\sum_ln_l$ with $n_l=p_l|\Psi_l^{pp}|^2$ as density
of electron orbital $l$. The potential $V_{\rm cor}$ given in eq.~(\ref{eq2}) aims 
at cancelling the apparent self interaction within orbital $l$~\cite{per81}.
The weight factors $w_l$ between 0 and 1 take care of adjusting the SIC correction to the 
crystal environment in view of the orbital occupations. Here, while the O$(2s)$ orbital 
is by default fully corrected with $w_{2s}=1.0$, the reasonable 
choice~\cite{kor10,lec19} $w_{2p}=0.8$ 
is used for the O$(2p)$ orbitals.  The resulting SIC PP in eq.~(\ref{eq3}) is of 
Kleinman-Bylander operator form and includes an additional parameter $\alpha$. That
second parameter proves useful to take are of further screening properties from the
crystal environment. As in previous DFT+sicDMFT 
works~\cite{lec19,lec20-1,lec20-2} we used $\alpha=0.8$, i.e. $\alpha=w_{2p}$ holds as
a single SIC parameter.

Introducing SIC for oxygen hence invokes a recomputation, and 
effective localization~\cite{tem93}, of the O orbitals, resulting in a revised
$r$-dependent pseudopotential. Focussing on O$(2p)$, this leads to two key effects: 
the onsite level energy is modified, attributed to an effective $U_{pp}$, and the
binding properties to Ni$(3d)$ altered, attributed to an effective $U_{pd}$. As a
result, the $pd$-splitting is enlarged and O$(2p)$-hybridized dispersions are 
narrowed. Albeit still on an effective single-particle level, these corrections 
thus carry over some energy dependence, while the standard DFT+U scheme~\cite{ani91_2} 
via its Hartree-Fock structure promotes essentially a high-energy correction. Moreover, note 
that explicit symmetry breakings are not necessary to obtain the SIC effects.

The inclusion of explicit oxygen-based correlations beyond conventional DFT+DMFT for 
the NdNiO$_2$ system is believed necessary for mainly two reasons. First generally, 
in nickelates as late TM oxides the charge-transfer aspect is 
stronger as in early TM oxides, and therefore Coulomb interactions from O$(2p)$ will 
have a larger impact~\cite{zaa85,ima98,han11}. 
Second, while Ni-$d_{x^2-y^2}$ is strongly hybridized with O$(2p)$ in the 
infinite-layer structure, Ni-$d_{z^2}$ is less so since there is no apical 
oxygen (see Fig.~\ref{fig:orb}). Thus especially the Ni-$e_g$ $\{z^2,x^2-y^2\}$ 
orbitals of the DMFT-active Ni$(3d)$ shell are quite differently affected by 
O$(2p)$ and therefore it should be insufficient to treat the Coulomb interactions 
from oxygen only implicitly via a chemical-potential or Coulomb-on-Ni shift. 
In the Appendix~\ref{siccomp}, we provide a brief comparison between the DFT+sicDMFT 
and DFT+DMFT picture of pristine and doped NdNiO$_2$.

\subsection{Calculational settings\label{sec:calc}}
A $13\times 13\times 13$ k-point mesh is utilized for the basic NdNiO$_2$ unit cell. 
This mesh is reduced for the magnetically-ordered supercells as to display the same
k-point density for enabling total-energy comparisons. The plane-wave cutoff energy
is set to $E_{\rm cut}=16$\,Ry and local basis orbitals are introduced for
Nd$(5d)$, Ni$(3d)$ as well as O$(2s,2p)$.
The Nd$(4f)$ states are put in the pseudopotential frozen core, since they are
not decisive for the key physics of infinite-layer nickelates~\cite{zha20}. This
also means, that $f$-electron based magnetism is not considered in this work.
As in many other comparable compounds, due to the local nature of the $4f$ moments 
and their weak exchange coupling, magnetic ordering of Nd moments is believed 
to occur at very low temperatures. Note furthermore that the $j$-resolved
frozen occupation of the Nd$(4f)$ shell is here chosen with small resulting 
moment. The role of spin-orbit effects in the overall crystal calculations is 
neglected. Additional details for the computation of the magnetically ordered 
phases in this work will be given in section~\ref{sec:afm}.

To facilitate the description at finite doping, we here employ the 
virtual-crystal approximation (VCA) in the fully charge self-consistent scheme. 
This amounts to replace the Nd atom by a pseudo atom of nuclear (and electronic) 
charge ${\cal Z}=Z_{\rm Nd}-\delta$ in the pristine unit cell. As a
choice, the charge content of the Nd$(4f)$ shell is here not modified by this 
construction. It is then assumed that $\delta=x$ holds, i.e. the VCA doping 
$\delta$ mimics the true Sr doping $x$ of effectively replacing Nd$^{3+}$ by 
Sr$^{2+}$. As any theoretical doping description,
also the VCA misses certain effects, e.g. local-structural distortions or 
correct local charge states of the impurity. But it usually provides a faithful
description of the overall effects of charge doping. Note that in the previous
works Refs.~\onlinecite{lec20-1,lec20-2}, the other two standard options to 
describe doping, namely supercell~\cite{lec20-1} and chemical-potential 
shift~\cite{lec20-1} have been put into practise. Thus the present perspective
completes the conventional method spectrum in that regard.

The DMFT correlated subspace is governed by a full Slater Hamiltonian
applied to the Ni$(3d)$ projected-local orbitals~\cite{ama08}. The projection is 
performed on the $6+5+1=12$ Kohn-Sham (KS) states above the dominant 
O$(2s)$ bands, associated with O$(2p)$, Ni$(3d)$ and the self-doping band. 
A Hubbard $U=10$\,eV and a Hund exchange $J_{\rm H}=1$\,eV prove reasonable for 
this choice of the energy window in the given nickelate~\cite{lec20-1,lec20-2,lec19}.
The fully-localized-limit double-counting scheme~\cite{ani93} is applied.
Continuous-time quantum Monte Carlo in hybridzation expansion~\cite{wer06} as 
implemented in the TRIQS code~\cite{par15,set16} is utilized to handle the 
DMFT problem. The system temperature is set to $T=30$\,K 
in order to properly approach the competition/cooperation between quasiparticle 
(QP) formation, Mott criticality, Kondo physics and magnetic order in these 
challenging nickelates.
Up to $2\cdot 10^{9}$ Monte-Carlo sweeps are performed for the
respective final convergence steps.  A Matsubara mesh of 2049 frequencies is
used to properly account for the low-temperature regime.
The standard maximum-entropy~\cite{gub91} 
and Pad{\'e}~\cite{vid77} method are employed for the analytical 
continuation from Matsubara space onto the real-frequency axis. 
If not otherwise stated, all shown data is obtained within 
charge self-consistent DFT+sicDMFT.

\section{Paramagnetic system}
\subsection{Stoichiometric compound}
A basic characterization of the correlated electronic structure of paramagnetic (PM)
NdNiO$_2$ has been given in Ref.~\onlinecite{lec20-1} for ambient temperatures, 
and in Ref.~\onlinecite{lec20-2} for the lower-temperature regime. To set the stage,
we here recapitulate the main findings and discuss the different site/orbital 
contributions within the correlated regime in some more detail.
\begin{figure*}[t]
\includegraphics*[width=17.75cm]{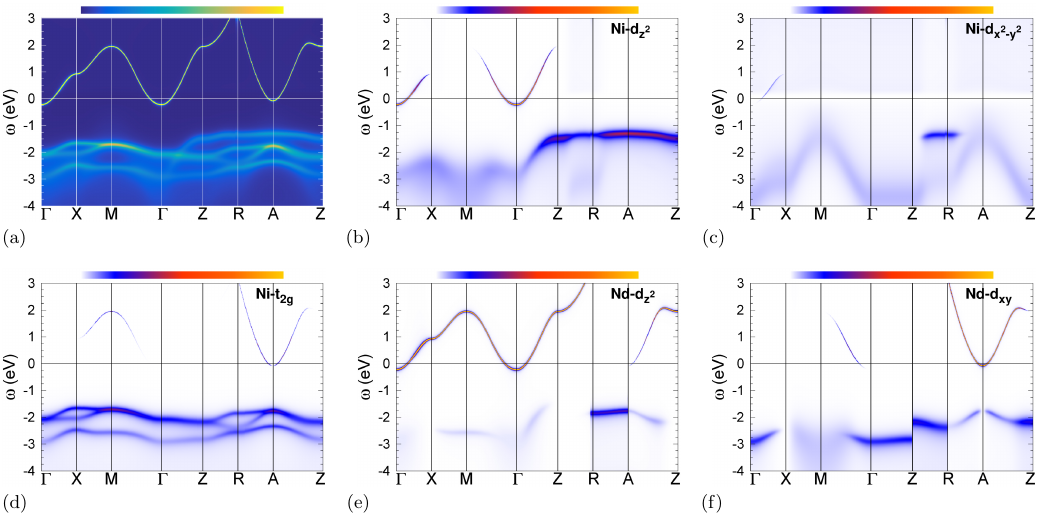}
\caption{(color online) Spectral information for pristine NdNiO$_2$.
(a) $\bk$-resolved function $A(\bk,\omega)$ along high-symmetry lines: $\Gamma$-X-M
describes a triangle in the $k_z=0$ plane and Z-R-A marks the same triangle translated
to the $k_z=1/2$ plane. Above the Fermi level, only the single SD dispersion is shown 
and further dispersions discarded in the plot.
(b-f) $\bk$-resolved orbital weights (i.e. fatbands) for the
given spectral function: (b) Ni-$d_{z^2}$, (c) Ni-$d_{x^2-y^2}$, (d) Ni-$t_{2g}$,
(e) Nd-$d_{z^2}$ and (f) Nd-$d_{xy}$. Note that with this chosen intensity resolution, 
the nondispersing-level feature at $\varepsilon_{\rm F}^{\hfill}$, giving rise to a 
Kondo(-lattice) mechanism~\cite{lec20-2} at lower $T$, is not visible in 
$A(\bk,\omega)$.}
\label{fig:stoich}
\end{figure*}
 \begin{figure}[b]
\includegraphics*[width=8.5cm]{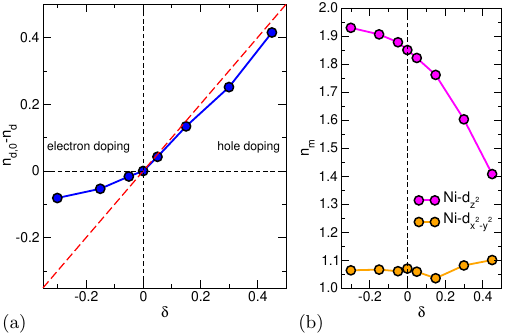}
\caption{(color online) Effective Ni$(3d)$ occupation in doped NdNiO$_2$ from 
introducing $\delta$ holes/electrons.
(a) Difference $n_{d,0}-n_d$ between the total occupation at $n_{d,0}$ zero 
doping and at given doping $n_d$, versus $\delta$. The dashed red line marks the
regime where the complete doping charge is transfered to Ni$(3d)$.
(b) Orbital-resolved filling of the Ni-$e_g$ states.}
\label{fig:occ}
\end{figure}

On the DFT(LDA) level, the $x^2$$-$$y^2$ orbital has the lowest crystal-field level 
within the Ni$(3d)$ sector and the energy splittings to the other levels read 
$\{\varepsilon_{z^2},\varepsilon_{x^2-y^2},\varepsilon_{xz},\varepsilon_{yz},
\varepsilon_{xy}\}=\{356,0,286,286,536\}$\,meV.
Then in essence, the compound is a self-doped Mott insulator in DFT+sicDMFT, whereby 
the effectively Mott-insulating states are of half-filled Ni-$d_{x^2-y^2}$ kind and 
the self doping originates from a weakly-occupied band resulting from the hybridization 
between Nd$(5d)$ and Ni$(3d)$. 
Thus there is no Ni-$d_{x^2-y^2}$ quasiparticle dispersion crossing the
Fermi level $\varepsilon_{\rm F}^{\hfill}$ in the $\bk$-resolved 
spectral function, as shown in Fig.~\ref{fig:stoich}a. At low energy, there are only 
SD-band based electron pockets at $\Gamma$ and A in the Brillouin zone.
\begin{figure*}[t]
\includegraphics*[width=17.75cm]{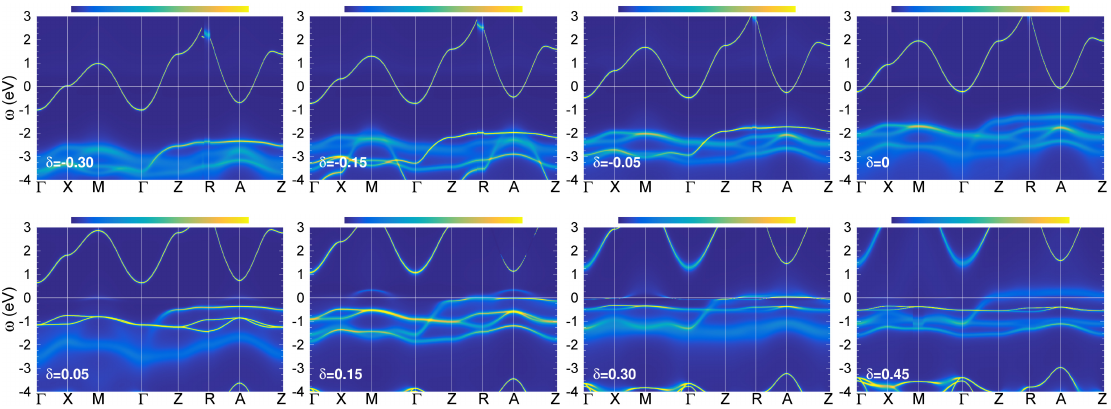}
\caption{(color online) Spectral function $A(\bk,\omega)$ with doping $\delta$.
Top row, from left to right: $\delta=-0.30,-0.15,-0.05,0$.
Bottom row, from left to right: $\delta=0.05,0.15,0.30,0.45$. Same setting as in
Fig.~\ref{fig:stoich}a.}
\label{fig:scan}
\end{figure*}

Correspondingly, orbital weights along the given high-symmetry lines are provided
in Fig.~\ref{fig:stoich}b-f for the most-relevant orbital sectors. The Ni-$d_{z^2}$
contribution is twofold. First, it forms an incoherent part around -3\,eV for 
$k_z=0$, which becomes more coherent, flat and intense in the $k_z=1/2$ plane. Second,
it has its prominent share in the SD-band formation for the $\Gamma$-pocket part. On the
other hand due to Mottness, the Ni-$d_{x^2-y^2}$ weight remains largely incoherent
troughout the chosen $\bk$-space regions, with the exception of a small part along
Z-R in the regime of the overall Ni-$d_{z^2}$-dominated spectrum. The 
Ni-$t_{2g}$ $\{xz,yz,xy\}$ orbitals cause dispersing features between [-3,-2]\,eV and
have weak contribution to the SD band around A. The Nd$(5d)$-based orbitals of
$z^2$ and $xy$ character are the main contributors to the SD band from the rare-earth
site; Nd-$d_{xy}$ for the A pocket and Nd-$d_{z^2}$ pretty much everywhere else.
The hybridization between Nd-$d_{xy}$ and Ni-$t_{2g}$ is also evident from the joint
weight at deeper energies. The coupling between Ni-$d_{z^2}$ and Nd-$d_{z^2}$ is
mostly confined to the $\Gamma$-X-M-Z path. Mott-insulating 
Ni-$d_{x^2-y^2}$ appears disconnected from the Nd$(5d)$-Ni$(3d)$ hybridization,
and indeed nearest-neighbor hopping from both relevant Nd$(5d)$ orbitals to the 
in-plane Ni-$e_g$ orbital is zero~\cite{nom19,lec20-1,lec20-2}. The O$(2p)$ contribution
is surely also vital in the given energy window, but a ${\bk}$-resolved discussion
appears not highly instructive (see Ref.~\onlinecite{lec20-1} for the $\bk$-integrated 
weight).

With high spectral-resolution focus on the low-energy region, a single-level feature 
at the Fermi level would become additionally visible for the here chosen lower 
temperature regime~\cite{lec20-2}. It is 
stemming from Ni-$d_{x^2-y^2}$ and is part of an intriguing Kondo(-lattice) mechanism,
involving furthermore Ni-$d_{z^2}$ and Nd-$d_{z^2}$. Basically, the Ni-$d_{z^2}$ 
orbital mediates a Hund-assisted Kondo coupling between the $d$ sectors of Ni and Nd,
by enabling fluctuations between Ni-$e_g$ and weakly-correlated Nd-$d_{z^2}$.

\subsection{Doped compound}
Superconductivity in the nickelate thin films is reached by Sr doping~\cite{li19}, 
expected to introduce holes to the system. For completeness we here discuss both, 
electron and hole doping within the VCA framework. From a preparation perspective, 
electron doping, e.g. by substituting Nd$^{3+}$ by Ce$^{4+}$, should be more 
challenging~\cite{hir19}: the already delicate Ni$^+$ ionic state is reduced even 
further and the charge imbalance within the NiO$_2$ plane becomes additionally 
stretched. Indeed, the DFT+sicDMFT calculations show that the progress of the doping 
level $\delta$ toward the Ni$(3d)$ shell is strongly hindered on the electron-doped
side (see Fig.~\ref{fig:occ}a). On the other hand on the hole-doped side, the doping
originating on the rare-earth site is nearly exclusively transfered to the Ni site.
This furthermore underlines the weaker charge-transfer character compared to cuprates,
where most holes reside on oxygen.
\begin{figure}[b]
\includegraphics*[width=8.5cm]{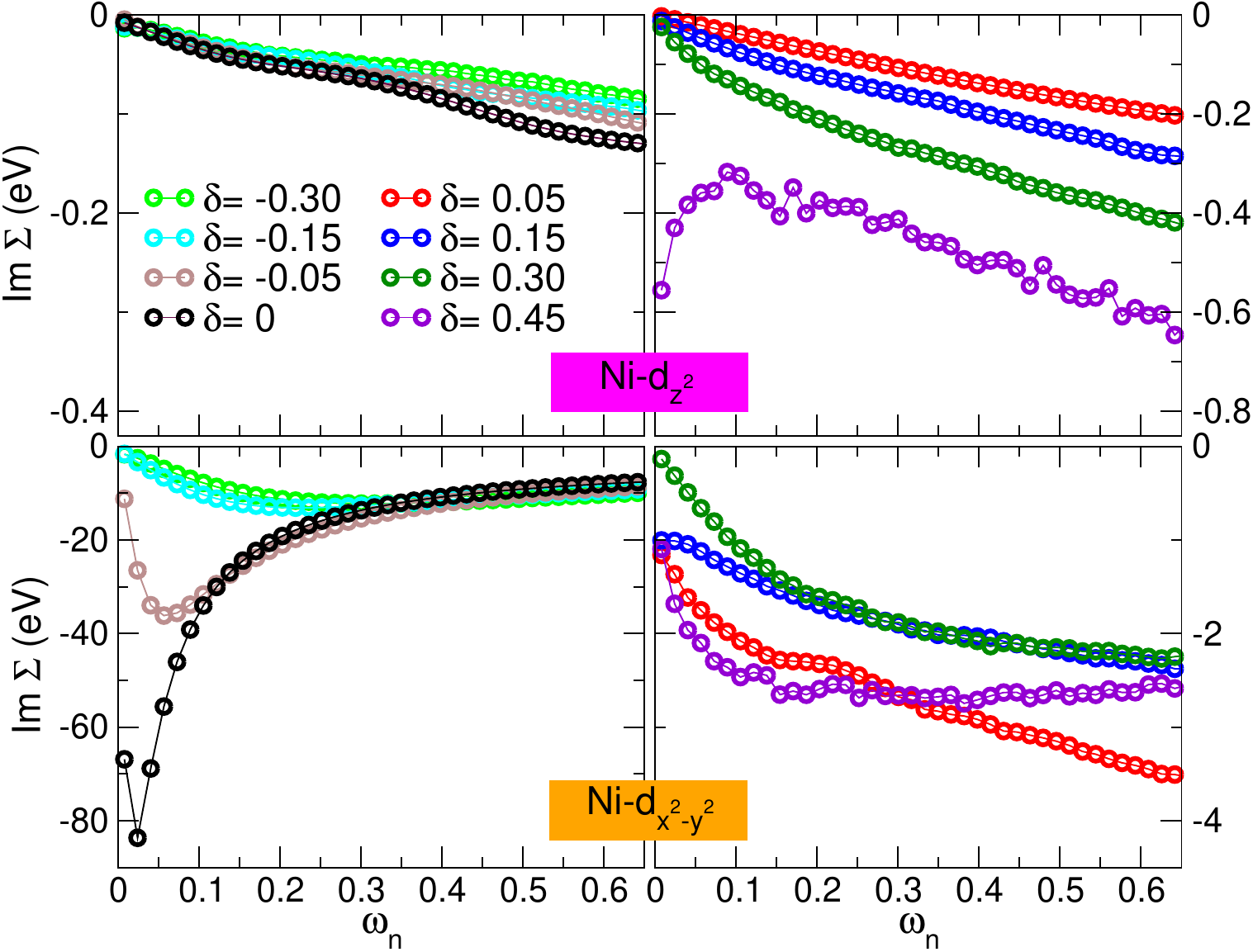}
\caption{(color online) Imaginary part ${\rm Im}\,\Sigma(i\omega_n)$ of the 
Ni-$e_g$ self energy on the Matsubara axis $\omega_n=(2n+1)\pi T$ for different
dopings $\delta$. Top: Ni-$d_{z^2}$ and bottom: Ni-$d_{x^2-y^2}$. Left part:
$\delta=-0.30,-0.15,-0.05,0$ and right part: $\delta=0.05,0.15,0.30,0.45$.}
\label{fig:sigscan}
\end{figure}
The orbital-resolved filling of the Ni-$e_g$ states with doping $\delta$ in 
Fig.~\ref{fig:occ}b exhibits the strong hole attraction of Ni-$d_{z^2}$ and its
weaker additional electron filling on the electron-doped side. On the contrary, 
the Ni-$d_{x^2-y^2}$ filling is only weakly affected by the charge doping of the
system and remains nearly inert. All these observations are in line with previous 
findings in a minimal-Hamiltonian description~\cite{lec20-2}. Note that the Ni-$t_{2g}$
states are always close to complete filling throughout the doping landscape and
seemingly do not play a key role for the phase diagram with $\delta$.
\begin{figure*}[t]
\includegraphics*[width=17.75cm]{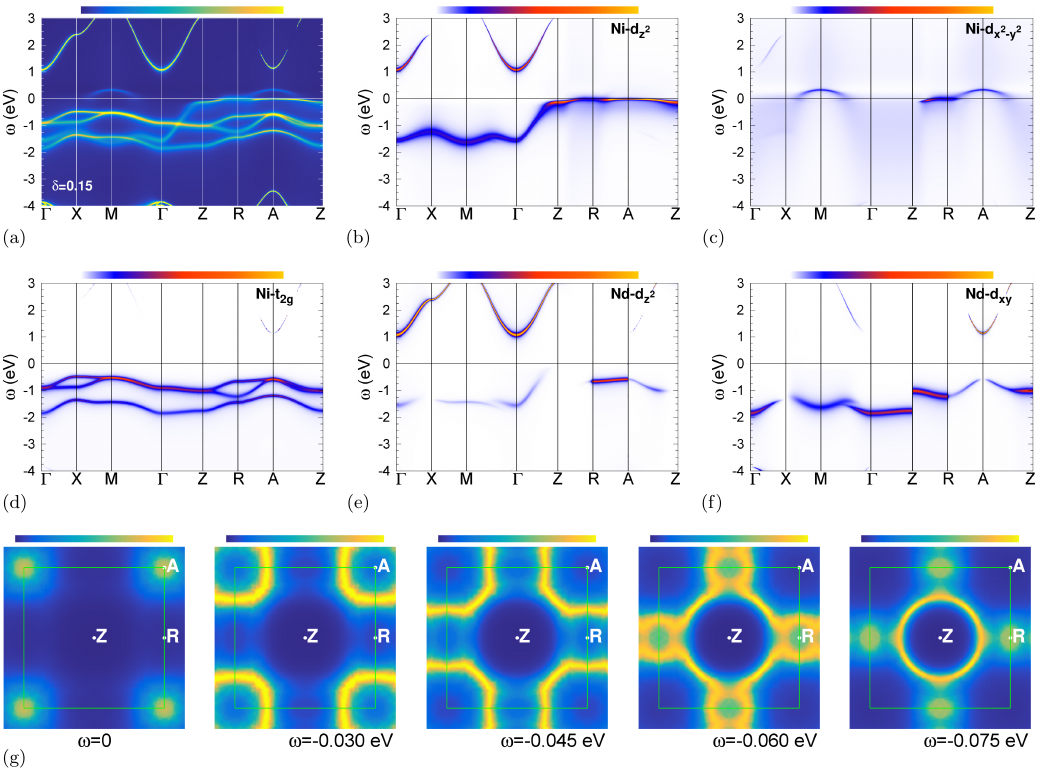}
\caption{(color online) Spectral information of hole-doped NdNiO$_2$ with $\delta=0.15$. 
(a-f) As in Figs.~\ref{fig:stoich}a-f. (g) Constant-energy surfaces in the $k_z=1/2$
plane. Note that the maximum intensity for the surfaces in (g) is reduced by a factor 
of three compared to the one for $A(\bk,\omega)$ in (a).}
\label{fig:d-akw}
\end{figure*}

We now turn to the ${\bk}$-resolved spectral evolution with doping, displayed in
Fig.~\ref{fig:scan}. Electron doping mainly leads to an expected downshift of the
overall spectrum, resulting in an increased filling of the SD band. The Mott-insulating
character of the Ni-$d_{x^2-y^2}$ orbital remains apparently robust, though the
corresponding self-energy is considerably weakened with electron doping 
(cf. lower-left part of Fig.~\ref{fig:sigscan}). Note however that this weakening
proceeds even stronger with hole doping (cf. lower-right part of 
Fig.~\ref{fig:sigscan}), as observed already in the minimal-Hamiltonian 
picture~\cite{lec20-2}. It leads to a partial melting of the Ni-$d_{x^2-y^2}$-based
Mott state for $\delta\geq 0.05$. Moreover, the SD electron pockets are shifted into the 
unoccupied region by sizable amount already for $\delta=0.05$. In line with this, the 
dominant Ni-$d_{z^2}$ spectral part in the occupied region gets shifted progressively 
toward the Fermi level. The spectral evolution of the originally occupied 
Ni-$d_{z^2}$ part is in fact one striking feature with hole doping. Its flat-band 
part in the $k_z=1/2$ plane crosses $\varepsilon_{\rm F}^{\hfill}$ here for 
$\delta\geq 0.15$.

Let us therefore pause with the further evolution to higher hole dopings for a moment 
and focus on the spectrum at $\delta=0.15$ (cf. Fig.~\ref{fig:d-akw}). The 
Ni-$d_{z^2}$ flat-band part is touching the Fermi level~\cite{lec20-2,pet20} 
and the strength of the Ni-$d_{x^2-y^2}$ Mott state is weakened, giving rise to  
minor dispersive spectral weight at low energy. The latter is strongest close to
the R point, where a van-Hove singularity settles. The occupied Ni-$t_{2g}$, Nd-$d_{z^2}$
and Nd-$d_{xy}$ weights appear generally strenghthened compared to the stoichiometric
case. They are also shifted to lower energy, but do not participate on the singular 
fermiology. The actual Fermi surface with obvious Ni-$d_{z^2}$ weight at the A point
is shown on the left of Fig.~\ref{fig:d-akw}g. Because of the flat-band character,
the topology of the constant-energy surface changes quite significantly within 
a $\sim 100$\,meV range below $\varepsilon_{\rm F}^{\hfill}$. Going down in energy from
$\omega=0$, a hole surface around A transforms into an electron surface around Z. The
substantial hole character just below the Fermi level is in line with experimental
findings of hole-like transport in that doping region~\cite{li20,zen20}. But note that
the ``actual'' Fermi surface for a given $0.125\lesssim \delta\lesssim 0.25$ in the
region with low-$T$ superconductivity is still very delicate, and perfect agreement 
between experiment and theory will be hard to reach. Nonetheless, we predict that the 
experimental $\delta=0.15$ Fermi surface has the appearance of one of the three 
$\omega=[0,-0.045]$\,eV plots of Fig.~\ref{fig:d-akw}g.

The Ni-$e_g$ self-energies have a rather intriguing development with hole doping
(cf. right part of Fig.~\ref{fig:sigscan}). 
For Ni-$d_{z^2}$, the self-energy part ${\rm Im}\,\Sigma(i\omega_n)$ grows with 
$\delta$, and the low-frequency curvature increasingly departs from standard 
Fermi-liquid behavior.
Fitting to the function $F=A\,\omega_n^{\alpha}$, the values $\alpha_\delta$ read
$\{\alpha_{0.05},\alpha_{0.15},\alpha_{0.40}\}=\{0.89,0.74,0.60\}$, while $\alpha=1$
holds in a Fermi liquid. Furthermore for $\delta=0.45$, the self-energy attains a
negative curvature, signalling strong bad-metal/insulating tendencies.
\begin{figure}[t]
\includegraphics*[width=8.5cm]{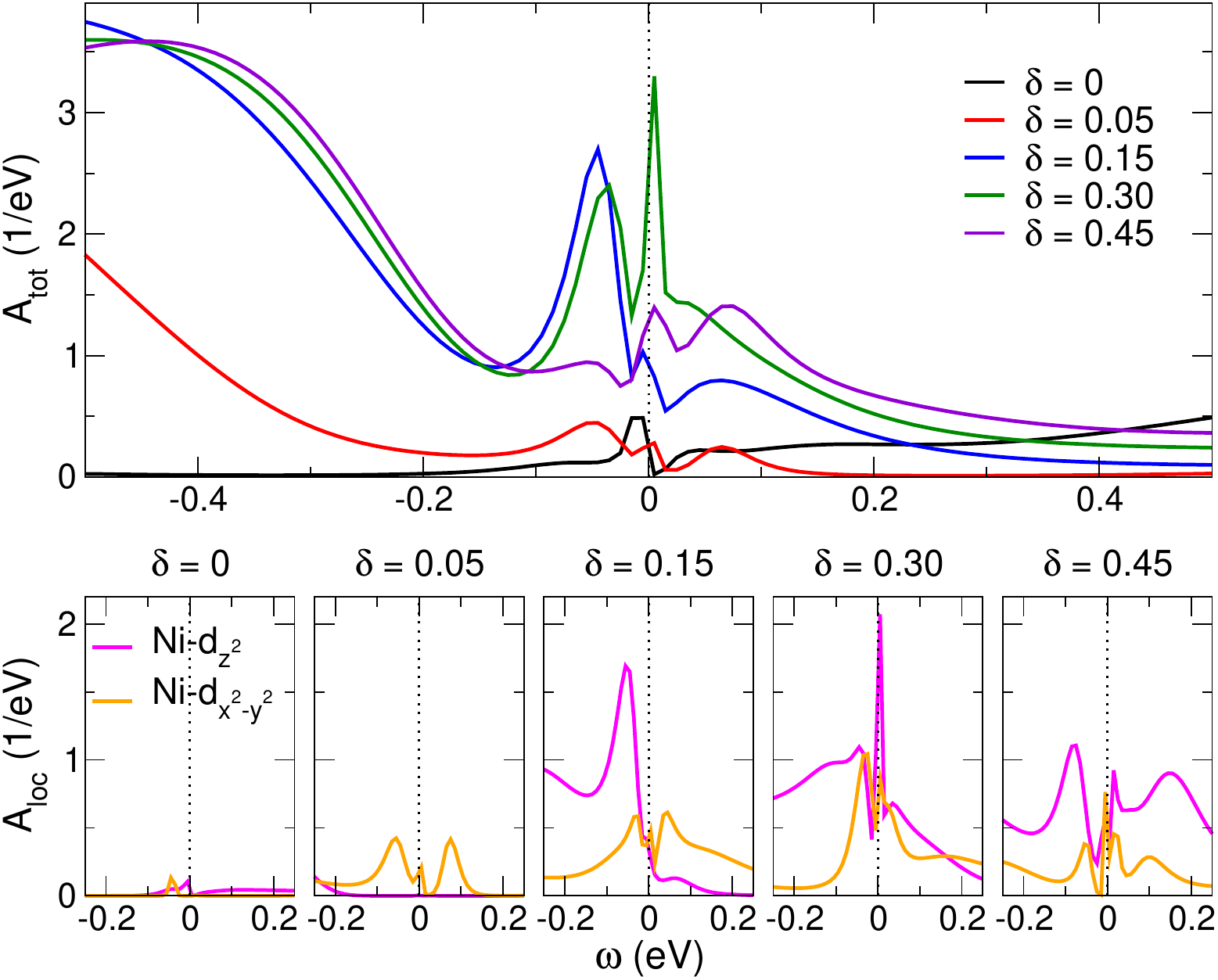}
\caption{(color online) Total spectral function after {\bf k}-integration (top) and 
local Ni-$e_g$ spectrum (bottom) of pristine and
hole-doped NdNiO$_2$ at low energy.}
\label{fig:shdop}
\end{figure}
A clearly larger Ni-$d_{x^2-y^2}$ self-energy shows nearly zero curvature at 
$\delta=0.15$, i.e. just when the Ni-$d_{z^2}$ flat band starts to cross 
$\varepsilon_{\rm F}^{\hfill}$. Increasing hole doping further, first stabilizes a
putative Fermi-liquid character with a low scattering rate 
($\propto {\rm Im}\,\Sigma(0)$), but then jumps back to a high-scattering Mott-critical
regime at $\delta=0.45$. This data shows that both Ni-$e_g$ orbitals
strongly affect each other in the hole-doped regime of NdNiO$_2$.

The development with hole doping can also be inspected from plotting the $\bk$-integrated
total spectrum and the local Ni-$e_g$ spectrum at low energy in Fig.~\ref{fig:shdop}. 
The total spectrum around $\varepsilon_{\rm F}^{\hfill}$ grows significantly for 
$\delta\geq 0.15$, but weakens again for $\delta=0.45$. The low-energy Ni-$e_g$ 
weight is small at stoichiometry; Ni-$d_{x^2-y^2}$ is essentially Mott insulating and
Ni-$d_{z^2}$ only weakly contributing based on its hybridization on the SD electron
pockets. Those are shifted above the Fermi level for small hole doping, and hence only
Ni-$d_{x^2-y^2}$ weight from a pre-melted Mott state appears close to 
$\varepsilon_{\rm F}^{\hfill}$. Once the flat band shifts into the low-energy
region, the corresponding Ni-$d_{z^2}$ weight starts to dominate the local spectrum.
But it develops a pseudogap structure due to specific dispersion topology of a
fully occupied(empty) flat branch below(above) the Fermi level. Note that the
increased QP character of Ni-$d_{x^2-y^2}$ between entry and departure of the Ni-$d_{z^2}$
flat-band part from the Fermi level becomes here also visible. As already derived from
the self-energies, the high-doping region of the hole-doped side does not resemble 
a good metal. But its semimetallic/bad-metallic appearance is indeed in agreement with
the weakly-insulating behavior found in experiment~\cite{li20,zen20}.

The filling scenario and $\bk$-space topology behind the Ni-$d_{z^2}$ pseudogap 
structure for hole dopings beyond the superconducting region is one reason 
for weak metallicity, especially for in-plane transport, as already pointed out in 
Ref.~\onlinecite{lec20-2}. However in addition, the non-Fermi-liquid and 
heavily bad-metallic character of the Ni-$d_{z^2}$-dominated states, 
associated with the pseudogap structure and a specific Hund-metal scenario 
of Ni-$e_g$, is a second vital reason.
\begin{table}[b]
\begin{ruledtabular}
\begin{tabular}{c|r r}
magnetic order & \multicolumn{2}{c}{$m_{\rm Ni}\,(\mu_{\rm B})$\;,\;
                 $E_{\rm tot}^{\rm AFM}-E_{\rm tot}^{\rm PM}$ (meV/atom)} \\[0.1cm]
               & $\delta=0$ & $\delta=0.15$ \\ \hline\\[-0.1cm]
G-AFM & 0.076\;,\;105 &  0.059\;,\;126  \\[0.1cm]
C-AFM & 0.081\;,\;20 &  0.000\;,\;$-$  \\[0.1cm]
\end{tabular}
\end{ruledtabular}
\caption{Ordered Ni magnetic moment in the G-AFM and C-AFM phase at $T=30$\,K and
energy difference to the PM phase.}
\label{tab:nimom}
\end{table}

\begin{figure*}[t]
\includegraphics*[width=17.75cm]{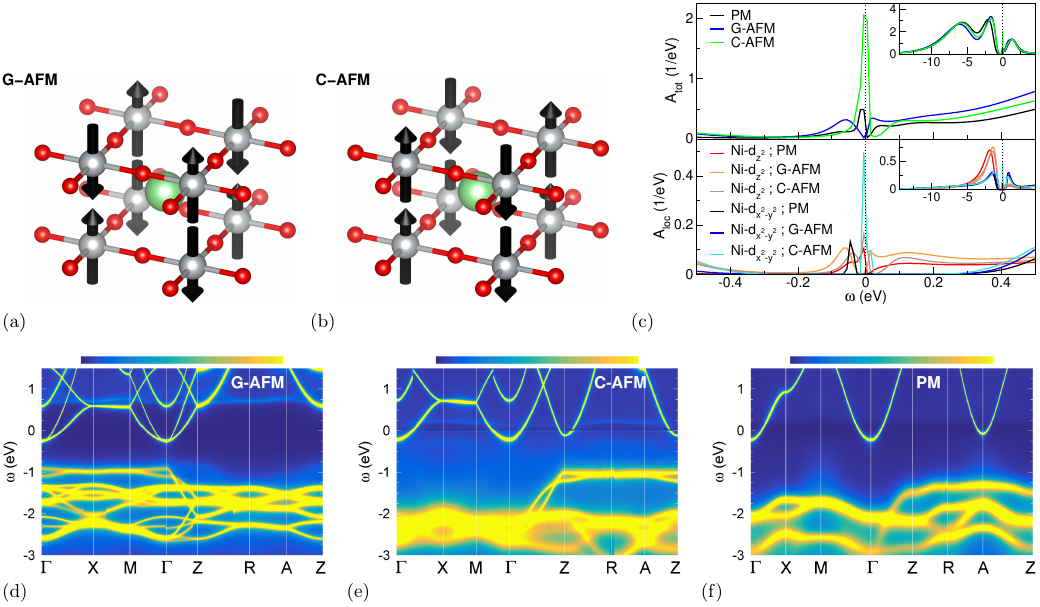}
\caption{(color online) Comparing PM and AFM-ordered NdNiO$_2$. (a) G-AFM ordering and (b)
C-AFM ordering. (c) Total (top) and local Ni-$e_g$ (bottom) spectral function
for PM, G-AFM and C-AFM phase, respectively. (d-f) $\bk$-resolved spectral function
along high symmetry lines for G-AFM (d), C-AFM (e) and PM (f) phase. The 
$\sqrt{2}\times\sqrt{2}$ enlarged in-plane unit cell translates into X$\leftrightarrow$M
and R$\leftrightarrow$A for the (G,C)-AFM structure. And the upper Ni-$d_{z^2}$ 
flat-band part is flipped with respect to $k_z$ for the G-AFM structure, due to the 
doubled unit cell along $c$.}
\label{fig:mspec}
\end{figure*}
\section{Antiferromagnetic system\label{sec:afm}}
A key feature of structurally akin high-$T_{\rm c}$ cuprates is the prominent AFM phase 
at and near stoichiometry~\cite{kei92,dag94}. So far, static magnetic order in 
infinite-layer nickelates has been elusive in experiment, but the investigations point
to a low-temperature behavior that deviates from straightforward 
paramagnetism~\cite{hay03,wang20,cui20}. 

From theory, there are various 
studies~\cite{lee04,hep20,bot20,cho20,zha20-2,gu19,liu20,leo20,leo20-2,zhalan20,kap20,cho20-2} 
that address ordered magnetism in these systems. Besides more unlikely ferromagnetic (FM)
order, the canonical orderings examined are G-type and C-type antiferromagnetism.
While the G-AFM order displays antiparallel alignment between in-plane and
$c$-axis nearest-neighbor (NN) Ni spins, the C-AFM order deviates therefrom in showing
ferro-alignment along the $c$-axis (see Figs.~\ref{fig:mspec}a,b). Most results from 
DFT(+U) and model-Hamiltonian DMFT calculations reveal a large ordered magnetic moment 
of $m_{\rm Ni}\sim 1$\,$\mu_{\rm B}$ on the Ni site for these patterns. 
At stoichiometry, the C-AFM state appears energetically favorable against the 
nonmagnetic state and against FM order within DFT+U~\cite{cho20-2} and additionally 
against G-AFM when utilizing the SCAN (i.e. meta generalized-gradient approximation)
functional in DFT~\cite{zhalan20}. However note that electronic correlations are only 
treated on a static level in these approaches (e.g. there is no true PM state 
straightforwardly reachable), and furthermore they formally describe the system at $T=0$. 
Nearest-neighbor in-plane exchange constants between Ni spins are estimated 
65\,meV~\cite{zhalan20} and 82\,meV~\cite{leo20-2}, i.e. roughly half the value 
usually revealed for high-$T_{\rm c}$ cuprates.

Here we report DFT+sicDMFT calculations for possible G-AFM and C-AFM ordering at 
$T=30$\,K. Two technical aspects have to be mentioned. First, the possible additional
ordering of the Nd spins is not considered in this investigation. This is not 
unreasonable, as in most rare-earth compounds the corresponding $4f$-based spins
order at much lower temperature. Second, in a hybrid scheme of DFT and DMFT, there
are in principle two basic formalisms to handle spin polarization: either it is
treated both in the DFT {\sl and} the DMFT part, or only on the DMFT level. The former
approach includes ligand-based exchange mechanisms more directly, but the 
$T$-independent exchange splitting from the DFT part usually ruins a reliable 
assessment of temperature effects. In other words, especially for transition-metal
oxides the robustness of the ordered magnetic moments with temperature is most often
severely overestimated. Therefore, spin polarization is here treated only on the 
DMFT level, i.e. the Ni$(3d)$ self-energies beyond DFT are solely active in 
creating Ni ordered moments in the charge self-consistent computations. Note however,
that strictly speaking, the leaking of the Ni projected-local orbitals toward
oxygen (cf. Fig.~\ref{fig:orb}) gives rise to spin-polarization effects beyond
pure Ni contributions.
The magnetic order is initialized by starting from the converged PM state and
using larger local magnetic fields on the given Ni sublattices in the first 
DFT+sicDMFT iteration.

The calculations are quite demanding and we so far did not succeed to
stabilize a magnetically ordered phase in electron-doped NdNiO$_2$. At stoichiometry
and with hole doping where this proves possible, the Ni magnetic moments turn out 
generally very small, with values $m_{\rm Ni}\lesssim 0.1$\,$\mu_{\rm B}$ 
(see Tab.~\ref{tab:nimom}). But importantly, the differentiation between an actual 
'zero moment' and such a small one is always very manageable from the present degree
of convergence and accuracy. Though surprising in view of the Mott-insulating 
Ni-$d_{x^2-y^2}$ orbital with formally sizable local $S=1/2$ spin at stoichiometry, 
Hayward and coworkers~\cite{hay03,hay99} noted from neutron scattering on 
polycrystalline bulk samples strong deviation from conventional $S=1/2$ local-moment 
behavior of Curie-Weiss form. In Ref.~\onlinecite{hay03}, the data analysis is
performed by assuming G-AFM order with a sensitivity limit of 0.06\,$\mu_{\rm B}$/Ni. 
\begin{figure}[t]
\includegraphics*[width=7.5cm]{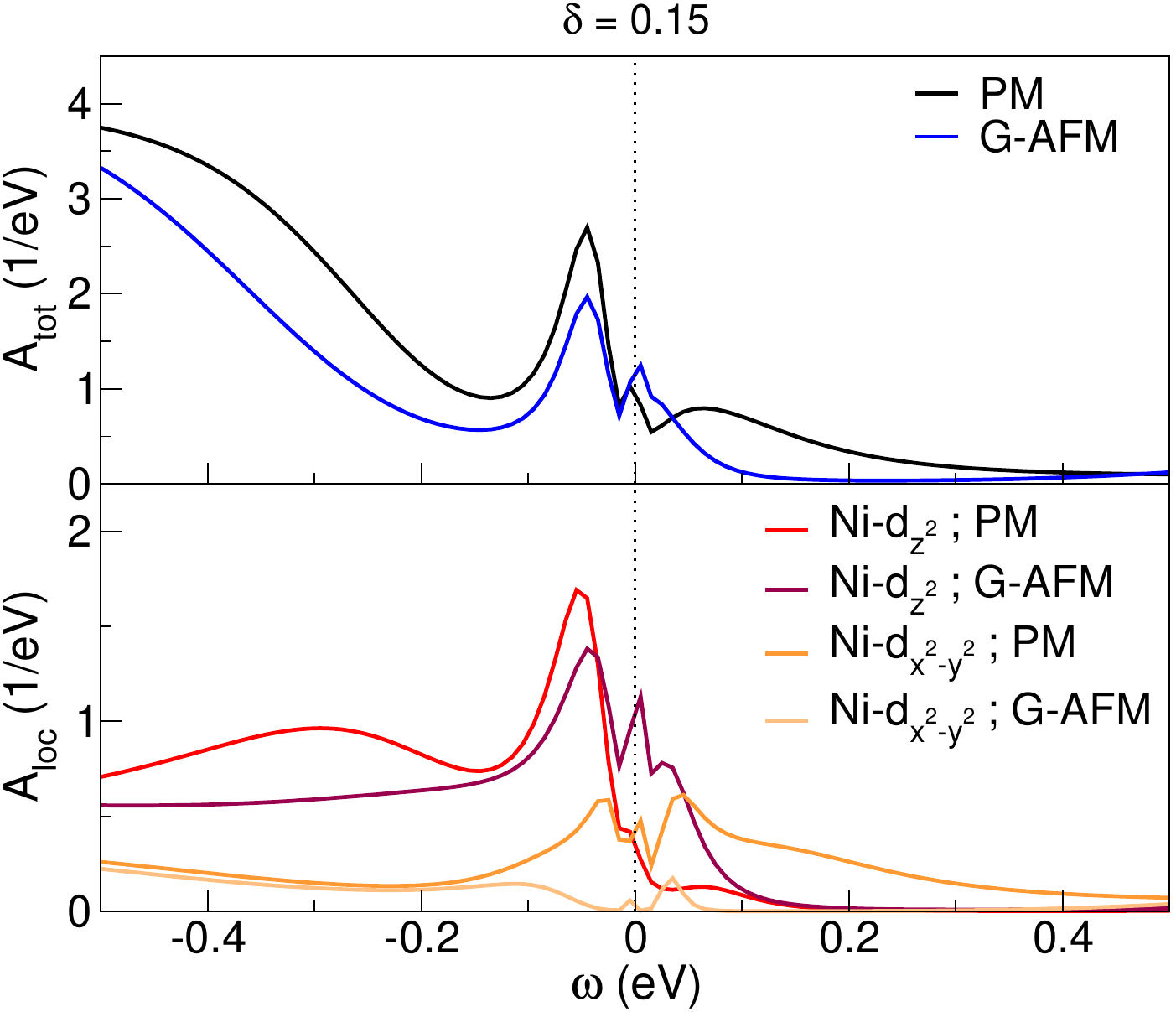}
\caption{(color online) Comparison of PM and G-AFM low-energy spectral functions 
at doping level $\delta=0.15$. (Top) Total spectrum and (bottom) local Ni-$e_g$ spectrum.}
\label{fig:dopmag}
\end{figure}

In the present investigation, supercells of $\sqrt{2}\times\sqrt{2}\times 1$ kind
for the C-AFM and of $\sqrt{2}\times\sqrt{2}\times 2$ for the G-AFM structure are
utilized. At stoichiometry, the PM phase has the lowest total energy for the given 
temperature.
But the energy difference $\Delta E=20$\,meV/atom to the C-AFM phase is rather small,
whereas the G-AFM phase with $\Delta E=105$\,meV/atom does not appear as a relevant 
competitor. For these three different phases, Fig.~\ref{fig:mspec}c displays total and 
local Ni-$e_g$ spectral functions, and Figs.~\ref{fig:mspec}d-f exhibit the
corresponding $\bk$-resolved spectra. Inspection of the total spectrum renders it 
obvious, that though the ordered moments are tiny, intriguingly, the low-energy spectra
of the different magnetic phases are actually quite different. Compared to the PM phase,
the small peak just below the Fermi level is substantially enhanced in the C-AFM phase
and more or less suppressed for the G-AFM phase. However, all three phases still display
the SD electron pockets at the Fermi level. Key difference between the G-AFM and C-AFM
low-energy physics is the enhancement of the Ni-$d_{x^2-y^2}$ single-level feature
at $\varepsilon_{\rm F}^{\hfill}$ for the latter phase. This becomes clear from the
cyan curve in the lower panel of Fig.~\ref{fig:mspec}c, as well as from the nondispersive
feature in $A(\bk,\omega)$ at the Fermi level in Fig.~\ref{fig:mspec}e. That 
Kondo(-lattice) resonance exists also in the PM phase~\cite{lec20-2}, yet with reduced
weight (and therefore not visible within the given resolution in Fig.~\ref{fig:mspec}e).
The C-AFM phase may thus be interpreted as an exotic phase with coexistence of 
antiferromagnetic ordering and Kondo(-lattice) screening~\cite{zha00,isa13,ber15,li15}.
Because of the weak filling of the the SD band, the latter Kondo physics takes place
in an underscreening scenario, leaving room for magnetic ordering. Additionally, since
the Kondo screening is mainly mediated by the $d_{z^2}$ orbitals of Ni and 
Nd~\cite{lec20-2}, the ferro-alignment of spins along the $c$-axis in the C-AFM phase 
enables a screening action without additional spin flip. This may explain as to
why the G-AFM phase cannot easily coexist with Kondo screening.

For finite hole doping $\delta=0.15$, the C-AFM phase has transformed
into the PM phase with zero ordered Ni moment. On the contrary, the G-AFM phase still
remains (meta)stable with somewhat reduced $m_{\rm Ni}$ compared to $\delta=0$
(see Tab.~\ref{tab:nimom}). Notably, as the SD pockets are shifted above the Fermi 
level for that hole-doping level, partial Kondo screening of the pristine-scenario 
kind has vanished. Figure~\ref{fig:dopmag} shows that close to the Fermi level, the 
Ni-$d_{z^2}$ character is enhanced and the Ni-$d_{x^2-y^2}$ one suppressed in the
doped G-AFM phase compared to the PM phase. Though energetically less favorable 
(see Tab.~\ref{tab:nimom}), if NN-AFM ordering is indeed stable at low temperature 
and with doping, there would be a $\delta$-induced transition from C-AFM to G-AFM 
ordering. 
Note that a qualitative change of magnetic order with doping has also been deduced 
by Leonov and Savrasov~\cite{leo20} from DFT+DMFT calculations for the system at 
ambient temperature, albeit differently with G-AFM order at stoichiometry and 
C-AFM order at finite hole doping.
Impact of method-based choices

\section{Impact of method-based choices}
Before summarizing and discussing the physical content of the obtained results, let
us weigh the influence of the performed choices within the DFT+sicDMFT scheme onto
the present findings.

Concerning the Coulomb interactions on Ni and O, the compatibility of the given 
Ni$(3d)$ onsite Hubbard $U$ and Hund's exchange $J_{\rm H}$ together with the present 
SIC configuration in view of describing the main NdNiO$_2$ transport characteristics 
has been already discussed in Ref.~\onlinecite{lec20-1}. The decisive impact of 
the SIC from the oxygen  sites in substantially enhancing the system correlation 
strength is reported in Appendix~\ref{siccomp}. A somewhat smaller $U$ value for 
Ni$(3d)$ of $U=7-9$\,eV would therefore not change the key findings by qualitative means. 
A smaller $J_{\rm H}$ would weaken the shifting of the Ni-$d_{z^2}$ flat-band part 
toward the Fermi level with hole doping, as the associated local Ni-$e_g$ $S=1$ state 
has to work harder against the crystal field~\cite{leo20-2}.
 
We did not explicitly employ local Coulomb interactions beyond the DFT treatment on the 
Nd site. First, such terms are believed significantly weaker than for Ni$(3d)$, and the 
Nd$(5d)$ filling is furthermore comparatively small. Second, the present choice of the 
correlated subspace with encompassing the 12 mainly Ni-O-dominated KS states as 
projection sphere, carries only minor explicit Nd$(5d)$ orbital weight. The body of
Nd$(5d)$ character is associated with higher lying KS states in the unoccupied part
of the spectrum~\cite{lec20-1}. To check the possible influence of Nd$(5d)$ interactions,
we therefore increased the KS projection sphere to 18 states and introduced an additional
local Slater Hamiltonian on the Nd site with $U_{\rm Nd}=3$\,eV and 
$J_{\rm H,{\rm Nd}}=0.3$\,eV solved within a now two-impurity DFT+sicDMFT cycle 
(while keeping interactions on Ni and SIC treatment for O unaltered). Note that this
enlarged number of KS states becomes then very similar to the projection window utilized
e.g. by Karp {\sl et al.}~\cite{kar20-1,kar20-2}. In this stronger atomic-orbital picture
of Ni$(3d)$ states, the Ni-$d_{z^2}$ turns out significantly {\sl lower filled} than in the 
stronger crystal-orbital 12-KS-state projection~\cite{kar21}, because of associated
orbital weight in the higher {\sl unoccupied} region~\cite{lec20-1}. The overall
physics at stoichiometry and with hole doping does not change qualitatively in 
DFT+sicDMFT employing the 18-KS-state projection. However already at stoichiometry, 
the occupied Ni-$d_{z^2}$ flat band in the $k_z=1/2$ part of reciprocal space is 
located much closer to the Fermi level. And for hole doping $\delta=0.15$, most of
this flat-band part has yet crossed $\varepsilon_{\rm F}^{\hfill}$. Note that within
the combination of DFT and DMFT, there is no unique criterion to design the correlated
subspace and thus to per se favor either of both KS projection spheres based on formal 
grounds. Based on our physical understanding, employing the 12-KS-state projection is
closer to the generic Hubbard-model construction idea, realizing a stronger Wannier(-like) 
picture for Ni$(3d)$, with a more stringent orbital-to-band resemblance. In addition,
high-lying KS states are less well-defined than occupied/low-lying ones, since the
Hohenberg-Kohn principles deal with the occupied charge density. Therefore, artifacts
or basis-dependent issues are more likely for high-energy unoccupied KS states.

Comparing to experiment, the here revealed behavior with hole doping does not seem 
perfectly commensurable with the experimental phase diagram. There, the superconducting 
region stretches from $0.125\lesssim \delta\lesssim 0.25$, whereas the prominent role of
Ni-$d_{z^2}$ seemingly peaks here at $\delta=0.30$. There may be several reasons for this
difference. First, due to our sparse doping sampling, a 'true' peaking Ni-$d_{z^2}$
behavior within the experimental superconducting doping range is still possible, as
for $\delta=0.30$ there is already the pseudogap onset. Second, we only assume $\delta=x$
to hold, but since the present VCA approach has its limitations, a small shift of
behavior would not be surprising. Moreover throughout the study, the stoichiometric
lattice parameters are employed, and e.g. the $c$-axis increase with hole 
doping~\cite{li19} is not included in the calculations. And third, albeit
the Ni-$d_{z^2}$ flat band neighorhood to the Fermi level is believed to be the key 
driving force for superconductivity, its $\varepsilon_{\rm F}^{\hfill}$ entry and 
departure points do not necessarily have to agree exactly with the phase boundaries 
for superconductivity.

The hybrid treatment, selection and initialization of magnetic order allows for 
alternatives. A different initialization protocol, i.e. starting from a self-consistent
constrained-field or Hartree-Fock solution, could possibly lead to the 
(meta)stabilization of larger-moment phases. Long-range ordering wave vectors 
beyond C- and G-AFM might be (meta)stabilized in extended considerations, which is 
however beyond the scope of this work.

Finally, the present status of the DFT+(sic)DMFT method asks for certain choices
and the final outcome does not only depend on pure convergence parameters as in
nowadays practical DFT. This may be seen as a drawback, but it emphasizes the complexity
of the interacting many-body problem and leaves room for physical intuition, which
has ever been a key aspect when pushing the frontiers of condensed matter.

\section{Summary and discussion}
The present DFT+sicDMFT study of pristine and doped NdNiO$_2$ uncovers two key results.
First, the character of the paramagnetic electronic structure changes distinctively
with doping. On the electron-doped side, which may be challenging to explore 
experimentally, the nickelate is a (Ni,Nd)-$d_{z^2}$ based one-band system in a 
Mott-critical background of Ni-$d_{x^2-y^2}$ kind. At stoichiometry, the latter 
background is of robust Mott-insulating nature and the solid one-band part at the
Fermi energy has evolved into a weakly-filled self-doping entity. Since the SD 
dispersion carries relevant Ni-$d_{z^2}$ itinerancy, one may also picture this phase 
as of orbital-selective type, eventually decorated with partial Kondo screening at 
low temperature~\cite{lec20-2}. Already small hole doping leads to a sizable upward 
shift of the SD band into the unoccupied region, and the stabilization of a 
Ni-$d_{x^2-y^2}$-based weakly-doped/pre-melted Mott-insulator in which Ni-$d_{z^2}$ 
degrees of freedom are frozen. 
For hole doping $\delta\gtrsim 0.15$, the Ni-$d_{z^2}$ flat-band 
part in the $k_z=1/2$ region crosses the Fermi level and the system displays 
intriguing $\{$Ni$-d_{z^2}$, Ni-$d_{x^2-y^2}\}$ physics at low energy. As a result 
thereof, unconventional superconductivity sets in at low temperature. For even larger 
hole doping $\delta>0.30$, the nickelate does not attain a conventional Fermi-liquid 
state, but enters a pseudogap and a bad Hund-metal regime. 
In other words, the two-orbital Ni-$e_g$ 
subspace with filling $n_{e_g}\sim 2.5$ becomes strongly correlated with vanishing 
coherence scale due to the interplay of the Hund's $J_{\rm H}$ with the remaining 
local Coulomb interactions in the Ni$(3d)$ and O$(2p)$ shells. Take notice that
other recent theoretical works mark general Nd$_{1-x}$Sr$_x$NiO$_2$ as a Hund metal, 
and suggest similarities with iron-pnictide physics~\cite{wankang20,kan20}.
However from our point of view, the (near) Mott-insulating state of Ni-$d_{x^2-y^2}$
and large filling of Ni-$d_{z^2}$ close to stoichiometry render this somewhat 
questionable. As described above, strong Hund-metal physics occurs here only 
for larger hole dopings, and therefore loosely speaking, superconductivity in 
doped NdNiO$_2$ appears to take place when 'cuprates' (from the underdoped side) and 
'pnictides' (from the overdoped side) meet and bind together through a 
flat-band scenario.

Second, it is shown that magnetic order of AFM type may be a vital competitor with and
without doping at low-enough temperatures, albeit the associated features are far from 
those of conventional antiferromagnetism in e.g. stoichiometric high-$T_{\rm c}$ 
cuprates. At $T=30$\,K, the ordered Ni moments are already very small for all 
encountered cases, which either points to a critical temperature in the same 
range or to strong quantum fluctuations/screening. The size of the local moment
surely also depends on calculational settings, e.g. on the method to handle spin-symmetry
breaking in realistic DMFT and the choice of the correlated subspace. But the present
choices are believed physically sound, and therefore the general outcome of 
significantly reduced Ni moments around $T=30$\,K should be qualitatively robust. 
The C-AFM phase is energetically more favorable than the G-AFM phase at stoichiometry, 
obviously benefiting from a coexistence with partial Kondo screening. This result 
sounds rather exotic for a TM($3d$) oxide, however the experimental searches for 
magnetic order in infinite-layer nickelates also point to singular low-$T$ 
behavior~\cite{hay03,wang20,cui20}. Furthermore, we here detect an apparent transition
from C-AFM to G-AFM order/correlations with hole doping. This would be coherent with
the fact that the Kondo-screening tendencies at stoichiometry vanish with hole doping,
because of the shift of the SD band above the Fermi level. 

This work complements our previous investigation for infinite-layer 
nickelates~\cite{lec20-1,lec20-2}. It is worthwhile to note that the obtained essential
picture of relevant Ni multiorbital physics, and especially the importance of the
Ni-$d_{z^2}$ flat-band part for hole dopings in the superconducting region, is very
robust across these studies. For instance, three different frameworks to describe the 
doped NdNiO$_2$ compound, i.e. supercell with impurities~\cite{lec20-1}, minimal
Hamiltonian with chemical-potential shift~\cite{lec20-2} and eventually the present
VCA approach, all lead to the same qualitative result. Thereby, the role of explicit 
Coulomb interactions on oxygen is not of 'decorative' type, but truly decisive for 
the revealed physics (see Appendix for further details). When abandoning the SIC 
inclusion into the theoretical framework and employing conventional DFT+DMFT 
calculations, the Ni-$d_{z^2}$ flat-band part remains well below the Fermi level for 
the relevant dopings, in agreement with comparable studies of this 
type~\cite{si20,leo20,kar20-2,wankang20}. Moreover, a recent GW+EDMFT investigation
of Petocchi {\sl et al.}~\cite{pet20}, which also includes beyond-Ni$(3d)$-based 
correlations, supports our picture.

Let us finally take the opportunity to selectively touch base with further
theoretical studies from other research groups. 
In DFT+U studies by Choi {\sl et al.}~\cite{cho20-2}, the C-AFM order also turns out
to be a proper candidate for magnetism in NdNiO$_2$. Furthermore, notably at 
stoichiometry, the Ni-$d_{z^2}$ flat-band part in the $k_z=1/2$ region is placed right 
at the Fermi level for this order within the given static-correlation scheme
(see also Appendix~\ref{app:u}). Though at a first glance looking quite different 
to our result for the C-AFM phase (compare with Fig.~\ref{fig:mspec}e), this outcome 
could carry similar physics. The DFT+U scheme may mimic Kondo(-like) physics by 
placing a spin-resolved flat band right at the Fermi level. In this way, the 
corresponding (local) spin becomes highly susceptible, and thats more or less 
the best a static-correlation scheme may achieve for Kondo-based spin fluctuations. 
The fact that this band is of dominant Ni-$d_{z^2}$ character underlines our previous 
finding of the crucial Ni-$d_{z^2}$ role in mediating Kondo-screening in 
NdNiO$_2$~\cite{lec20-2}. However, this DFT+U result for Ni-$d_{z^2}$ 
{\sl at stoichiometry} should not be directly related to the DFT+sicDMFT 
result of the Ni-$d_{z^2}$ flat-band part crossing the Fermi level 
with {\sl sizable hole doping}.

In a different DFT+U assessment by Liu {\sl et al.}~\cite{liu20}, the G-AFM order is
found to be stable at stoichiometry, but the Ni magnetic moment and the general AFM ordering
is also easily weakened by the impact of the SD band. The resistivity upturn below 
$T\sim 70$\,K is associated with the transition into this bad-AFM metal when lowering
temperature from the PM phase. Albeit details differ, that bad-AFM metal regime
could resemble our AFM plus Kondo-screening phase.

Finally, all our theoretical studies on the thin-film NdNiO$_2$ system, i.e. 
Refs.~\onlinecite{lec20-1,lec20-2} and this one, are conducted in a 'bulk-like' 
fashion by using the thin-film lattice parameter for a bulk-crystal calculation.
Already several works~\cite{ber20,gei20,he20,zhalin20} addressed the explicit 
thin-film geometry on a SrTiO$_3$ substrate via supercell DFT(+U) calculations. 
However, while nickelate/SrTiO$_3$ interface physics may have some impact, from the
comparison of the currently available experimental data and the results of a 
theoretical bulk-like approach there is no pressing evidence that key effects 
are missing in the latter. Furthermore, the experimental thin films with $\sim 10$\,nm
thickness are rather bulky, accounting for more than thirty vertical unit cells. 
The measured critical current densities~\cite{osa20-1} of more than 300 kA/cm$^2$ 
would also be quite large for sole twodimensional superconductivity in the interface. 
Nonetheless, investigating NdNiO$_2$ in an ultrathin limit of only a few unit cells 
would surely be interesting.

In the end, different theoretical concepts for infinite-layer nickelates are presently 
put to the test (see e.g. Ref.~\onlinecite{bot21} for a recent review). 
Ultimately, only experiment can tell which of these concepts 
is closest to the truth. The here presented evolution of the doping-dependent 
phenomenology is already in line with various experimental 
findings~\cite{li20,zen20,goo21}, most notably concerning the identification of 
weakly-insulating regions on either side of the superconducting doping region as well 
as the dominance of hole-like transport for $\delta>0.15$. In a next step, 
angle-resolved photoemission spectroscopy (ARPES) measurements appear deciding, since 
therefrom the correlation strength and the fate of the Ni-$d_{z^2}$ flat-band part with 
doping is most clearly revealed. But ARPES on these thin film architectures, together 
with the necessary resolution for possible flat-band physics out of the $k_z=0$ plane, 
is surely demanding.
By any means, the infinite-layer nickelate physics is highly puzzling and appears
to integrate various condensed-matter regimes into one materials class, providing
room for intense research in the future.
\begin{figure}[b]
\includegraphics*[width=8cm]{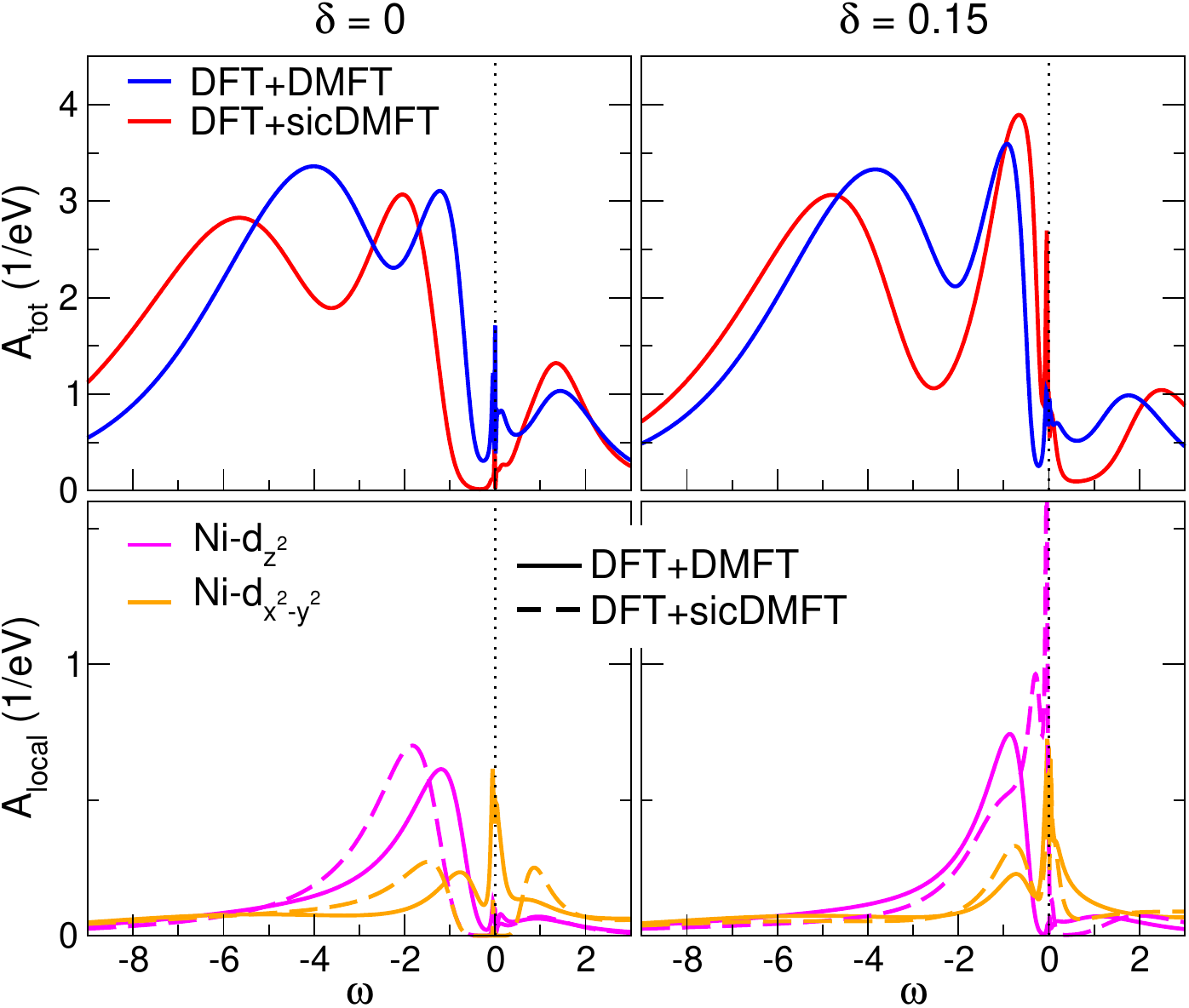}
\caption{(color online) Comparison of DFT+DMFT and DFT+sicDMFT spectral functions 
for pristine NdNiO$_2$ (left) and at doping level $\delta=0.15$ (right). (Top) 
Total spectrum and (bottom) local Ni-$e_g$ spectrum.}
\label{fig:sic2}
\end{figure}
\begin{acknowledgments}
The author thanks I. Eremin, A. Hampel, J. Karp, and A. J. Millis for 
helpful discussions. Financial support from the German Science Foundation (DFG) via 
the project LE-2446/4-1 is acknowledged. Computations were performed at the University 
of Hamburg and the JUWELS Cluster of the J\"ulich Supercomputing Centre (JSC) under project 
number hhh08.
\end{acknowledgments}

\appendix
\section{Role of explicit Coulomb interactions on oxygen\label{siccomp}}
As discussed in more detail in Ref.~\onlinecite{lec19}, the SIC 
treatment for most relevant O$(2p)$ in the present form effectively introduces a 
$U_{\rm pp}$ and a $U_{\rm pd}$ interaction term. As a result, the $pd$-splitting 
between O$(2p)$ and Ni$(3d)$ is enhanced compared to LDA, and the hoppings $t_{pd}$ 
between those site-orbitals are renormalized.
In order to assess the effect of including explicit Coulomb interactions on 
oxygen via a SIC-modified O pseudpotential, we performed also calculations within
the conventional DFT+DMFT scheme for the paramagnetic system. There, the standard 
O pseudpotential based on LDA is employed.
\begin{figure}[t]
\includegraphics*[width=8.5cm]{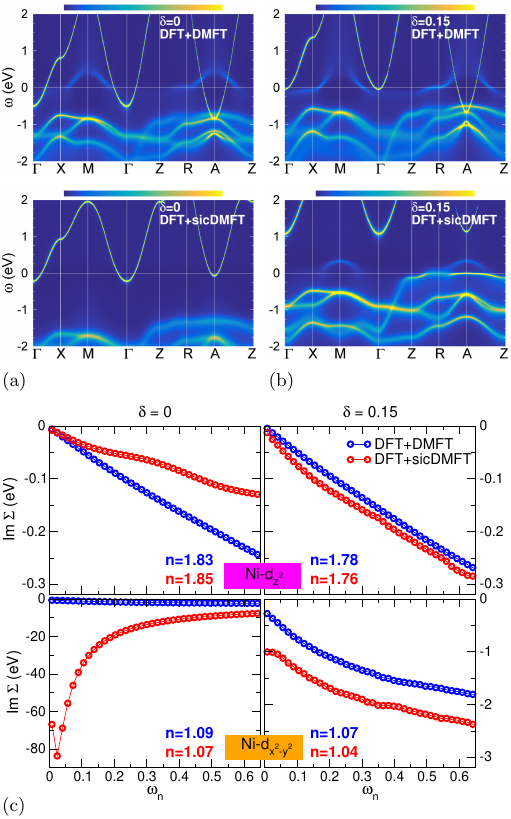}
\caption{(color online) Comparison between DFT+DMFT and DFT+sicDMFT for pristine (left)
and $\delta=0.15$ hole-doped (right) NdNiO$_2$. (a) $\bk$-resolved spectral function 
$A(\bk,\omega)$ and (b) imaginary part of the Ni-$e_g$ self-energy 
${\rm Im}\,\Sigma(i\omega_n)$.}
\label{fig:sic1}
\end{figure}

In Fig.~\ref{fig:sic2}, the total spectral function $A_{\rm tot}(\omega)$ and the
local Ni-$e_g$ part are shown in comparison for both computational schemes. One may
infer that though the Hubbard $U$ on the Ni site is identical, the overall
correlation strength in the standard DFT+DMFT approach is much weaker than with 
including additionally Coulomb interactions on oxygen. At stoichiometry, the system
is a strongly renormalized metal, i.e. with Ni-$d_{x^2-y^2}$ not reaching a
Mott-insulating state, in line with other works~\cite{si20,kar20-2,wankang20}. 
This can be explained by the fact that an effective treatment of $U_{pp}$ and 
$U_{pd}$ is missing in conventional DFT+DMFT. Thus a sole focus on the Hubbard $U$ 
on Ni in order to assess the correlation strength in infinite-layer nickelates may 
be misleading, since a vital contribution is carried by the Coulomb interactions on 
and from oxygen because of the late-TM-oxide context.
Furthermore, the distance between oxygen and Ni$(3d)$ states, associated with 
both main peaks in the respective total spectrum, is somewhat larger from the 
DFT+sicDMFT calculations, which reflects the increase of $pd$ splitting. 

For the hole-doped compound, it is additionally seen that the Ni-$d_{z^2}$ character 
does not show a peak close to the Fermi level for DFT+DMFT. This is understood from
Fig.~\ref{fig:sic1}, which displays the comparison between DFT+sicDMFT and DFT+DMFT 
on the level of the $\bk$-resolved spectral function $A(\bk,\omega)$ as well as the
imaginary part of the Ni-$e_g$ self-energy ${\rm Im}\,\Sigma(i\omega_n)$. 
It becomes obvious
that the previously discussed key effect of hole doping, namely the shift of the 
Ni-$d_{z^2}$ flat-band part in the $k_z=1/2$ plane is missing in DFT+DMFT. Moreover,
the SD electron pockets are much weaker shifted to higher (unoccupied) energies,
also in line with previous DFT+DMFT studies~\cite{si20,leo20,kar20-2,wankang20}. This
difference in doping evolution of those specific dispersions has to be 
related to the fact, that both individual Ni-$e_g$ orbitals hybridize very differently 
with O$(2p)$ (see last paragraph in section~\ref{sec:med}). Consequently, the 
Ni-$d_{x^2-y^2}$ self-energy at $\delta=0.15$ also does not show a clear low-frequency
anomaly in DFT+DMFT, since that one should be linked to the Ni-$d_{z^2}$ flat-band part
residing at $\varepsilon_{\rm F}^{\hfill}$. However, note that the Ni-$e_g$ occupations 
up to $\delta=0.15$ do not seem to differ strongly between both correlation schemes.

In summary, there are strong qualitative differences between the DFT+DMFT and the
DFT+sicDMFT description of pristine and doped NdNiO$_2$, which ask for a rethinking 
of the sole TM-driven correlation picture of late transition-metal oxides in many
numerical approaches.

\section{DFT+sicU result for the C-AFM phase\label{app:u}}
\begin{figure}[t]
\includegraphics*[width=8.5cm]{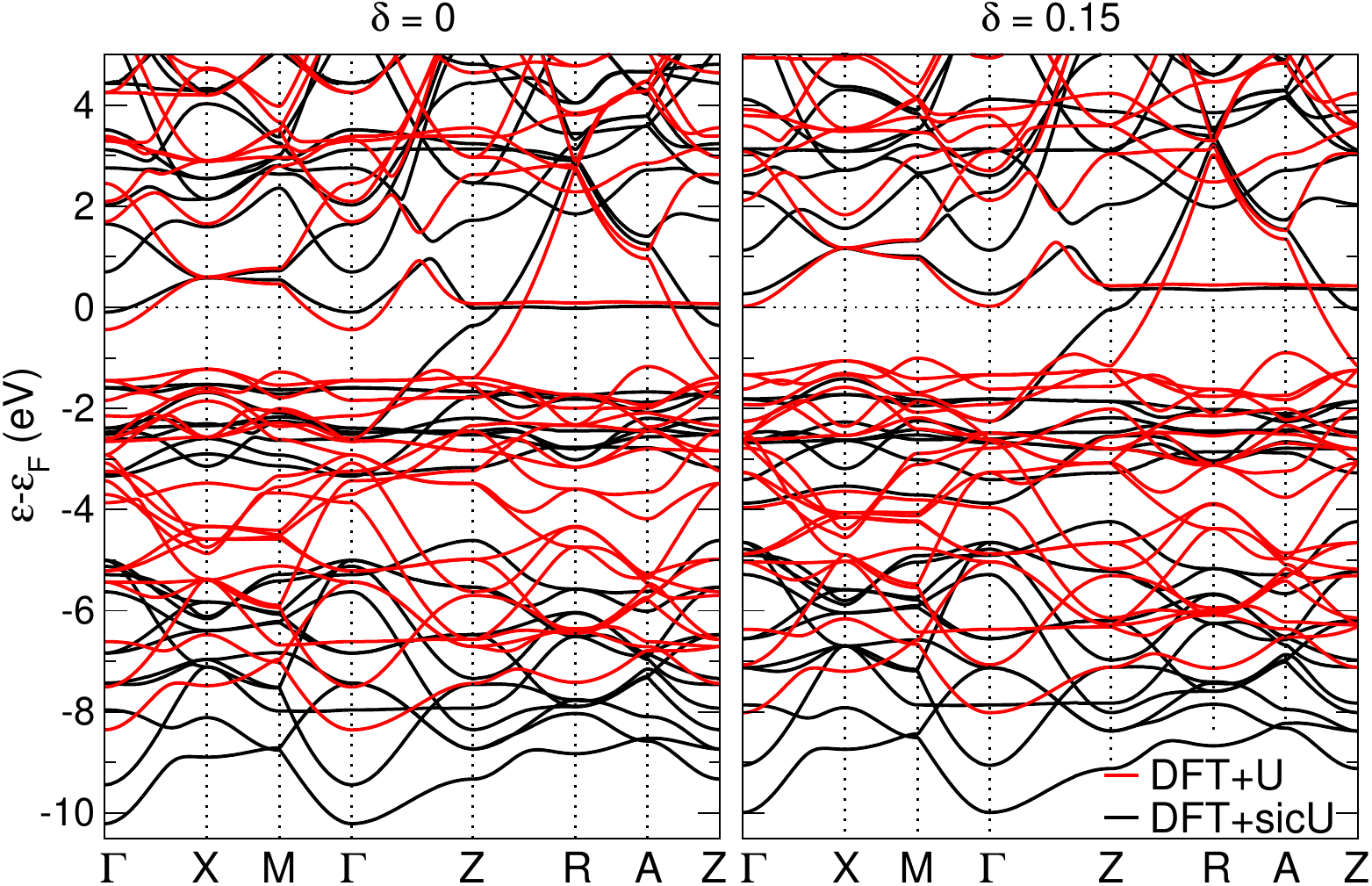}
\caption{(color online) Comparison between the DFT+U (red) and DFT+sicU 
band structure for pristine (left) and $\delta=0.15$ hole-doped (right) NdNiO$_2$
in the C-AFM phase.}
\label{fig:dftu}
\end{figure}

For comparison, we plot in Fig.~\ref{fig:dftu} the band structures obtained within 
DFT+U and DFT+sicU for the C-AFM phase. 
The DFT+sicU scheme amounts to the conventional DFT+U framework, but
here employing the SIC-modified oxygen pseudopotential. It hence mirrors the
DFT+sicDMFT scheme for a Hartree-Fock-like treatment of electronic correlations.
Note however that the correlated
subspace in this standard implementation of DFT+U is defined by projecting all KS
bands onto the Ni$(3d)$ cubic harmonics, as done in most available electronic structure
codes. A Hubbard $U=5$\,eV and $J_{\rm H}=1$\,eV is utilized in line with values 
used in other DFT+U studies of NdNiO$_2$~\cite{lee04,hep20,cho20-2}. No Hubbard 
interactions are used on the Nd site.

We confirm the flat-band behavior in the $k_z=1/2$ region already at stoichiometry, as
described in Refs.~\onlinecite{cho20-2,zhalan20}. This result holds with inclusion of
SIC on oxygen, but further renormalization of the electron pockets takes place close to
the Fermi level. The wide band crossing the Fermi level along Z$-$R is of dominant 
Ni-$d_{xz,yz}$ and Nd-$d_{xy}$ character.
At higher occupied energies, the $pd$ splitting is increased as expected
from the SIC inclusion~\cite{lec20-1}. Note that a sizable downward shift of the O$(2p)$ 
states is also obtained from $GW$ calculations~\cite{ole20}. With hole doping, the 
electron pockets become depleted and the flat-band part is also shifted above the Fermi 
level. The Ni AFM moments are sizable and read $m_{\rm Ni}=\pm 1.1(1.0)\,\mu_{\rm B}$ 
without(with) inclusion of SIC, and are even slightly growing in size with hole doping.

\bibliography{bibwrap}

\end{document}